\documentclass[12pt,a4]{article}
\usepackage{amsmath}
\usepackage{bm}
\usepackage{amsfonts}
\usepackage{amssymb}
\usepackage{graphics}
\usepackage[normal]{caption2}
\usepackage{subfigure}
\usepackage{rotating}
\usepackage{citesort}
\def\be{\begin{equation}}
\def\ee{\end{equation}}
\def\ba{\begin{array}}
\def\ea{\end{array}}
\def\bea{\begin{eqnarray}}
\def\eea{\end{eqnarray}}

\begin{document}
\baselineskip 20pt \setlength\tabcolsep{2.5mm}
\renewcommand\arraystretch{1.5}
\setlength{\abovecaptionskip}{0.1cm}
\setlength{\belowcaptionskip}{0.5cm}
\pagestyle{empty}
\newpage
\pagestyle{plain} \setcounter{page}{1} \setcounter{lofdepth}{2}
\begin{center} {\large\bf Study of participant-spectator matter and thermalization in isospin asymmetric systems}\\
\vspace*{0.4cm}

{\bf Sakshi Gautam} and {\bf Rajeev K. Puri}\footnote{Email:~rkpuri@pu.ac.in}\\
{\it  Department of Physics, Panjab University, Chandigarh -160
014, India.\\}
\end{center}

\section*{Introduction}
Besides the many radioactive ion beam facilities that already
exist in the world, a number of next-generation radioactive beam
facilities are being constructed or planned. At these facilities,
nuclear reactions involving nuclei with large neutron or proton
excess can be studied, thus providing a great opportunity to study
both the structure of rare isotopes and the properties of isospin
asymmetric nuclear matter. The ultimate goal of these studies is
to determine the isospin dependence of the in-medium nuclear
effective interactions and the equation of state of isospin
asymmetric nuclear matter. Role of isospin degree of freedom has
been investigated in disappearance of flow (at balance energy
(E$_{bal}$)) for the past decade. The first study predicting the
isospin effects in E$_{bal}$ was done by Pak \emph{et al}
\cite{pak}. Later on Liewen \emph{et al.} \cite{liewen} also
demonstrated isospin effects using IQMD in disappearance of flow.
The study revealed that the isospin effects in flow is due to the
competition between nucleon-nucleon collisions, symmetry energy,
surface properties and Coulomb force. But relative importance
among these mechanisms was not clear. One of the authors and
collaborators demonstrated the dominance of Coulomb potential in
isospin effects (for isobaric pairs) \cite{gaum1}. Motivated by
the dominance of Coulomb potential in isobaric pairs, authors and
collaborators studied flow in isotopic pairs and revealed its
sensitivity to symmetry energy in Fermi energy region
\cite{gaum2}. Hence many studies exist in literature which have
been carried out showing isospin effects in balance energy of
neutron-rich colliding pairs, but none of the study deals with
other heavy-ion phenomena at balance energy. Since at balance
energy the attractive mean field potential is balanced by
repulsive nucleon-nucleon interactions, so this counterbalancing
is reflected in quantities like participant-spectator matter etc.
In the present work, we study participant-spectator matter,
thermalization reached in the reactions of neutron-rich colliding
pairs at the energy equal to the the balance energy. The study is
carried out within the framework of IQMD model \cite{hart98}.

\section*{Results and discussion}
  We simulate the reactions of Ca+Ca, Ni+Ni, Zr+Zr, Sn+Sn, and Xe+Xe
series having N/Z = 1.0, 1.6 and 2.0. In particular, we simulate
the reactions of $^{40}$Ca+$^{40}$Ca (105), $^{52}$Ca+$^{52}$Ca
(85), $^{60}$Ca+$^{60}$Ca (73); $^{58}$Ni+$^{58}$Ni (98),
$^{72}$Ni+$^{72}$Ni (82), $^{84}$Ni+$^{84}$Ni (72);
$^{81}$Zr+$^{81}$Zr (86), $^{104}$Zr+$^{104}$Zr (74),
$^{120}$Zr+$^{120}$Zr (67); $^{100}$Sn+$^{100}$Sn (82),
$^{129}$Sn+$^{129}$Sn (72), $^{150}$Sn+$^{150}$Sn (64) and
$^{110}$Xe+$^{110}$Xe (76), $^{140}$Xe+$^{140}$Xe (68) and
$^{162}$Xe+$^{162}$Xe (61) at an impact parameter of
b/b$_{\textrm{max}}$ = 0.2-0.4 at the incident energies equal to
 balance energy. The values in the brackets represent
the balance energies for the systems.

\begin{figure}[!t] \centering
 \vskip -1cm
\includegraphics[width=7.5cm]{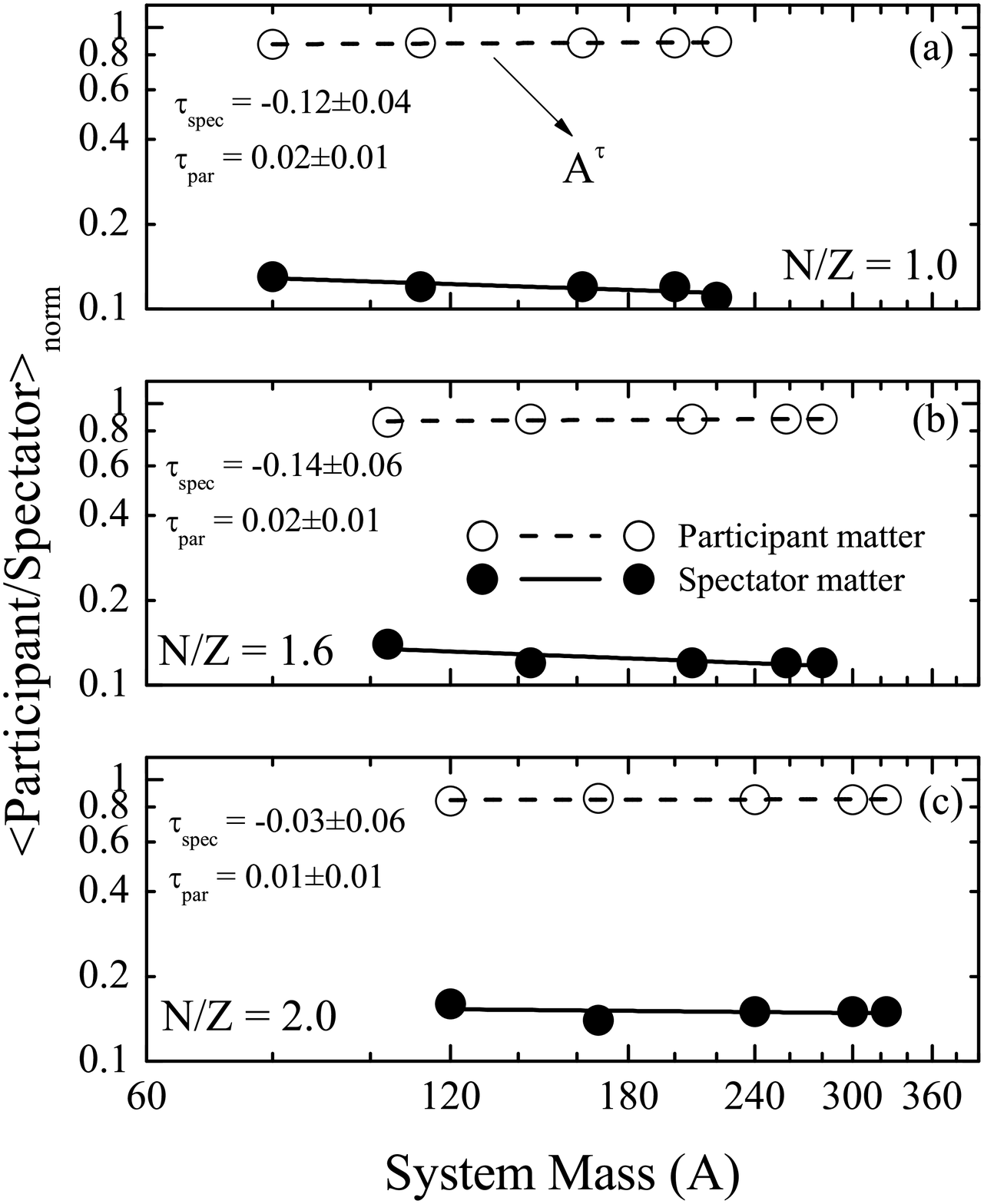}
\caption{The system size dependence of participant-spectator
matter for different N/Z ratios.}\label{fig1}
\end{figure}

\par
We define the participant-spectator matter in terms of nucleonic
concept, i.e, all those nucleons which suffer at least one
collision is called participant matter and the remaining nucleons
are termed as spectator matter. In fig. 1, we display the system
size dependence of the participant and spectator matter. Open
(solid) symbols represent participant (spectator) matter. Upper,
middle and lower panels represent the results for N/Z = 1.0, 1.6
and 2.0, respectively. We see that participant-spectator matter
follows a power law behaviour ($\propto$ A$^{\tau}$) with the
system mass. The power law factor is -0.12$\pm$ 0.04 (0.02$\pm$
0.01), -0.14$\pm$ 0.06 (0.02$\pm$ 0.01), and -0.03$\pm$ 0.06
(0.01$\pm$ 0.01) for spectator (participant) matter having N/Z =
1.0, 1.6 and 2.0, respectively. Thus, a nearly mass independent
behavior is obeyed by the participant and spectator matter for all
the N/Z ratios.

\begin{figure}[!t] \centering
 \vskip -1cm
\includegraphics[width=7.5cm]{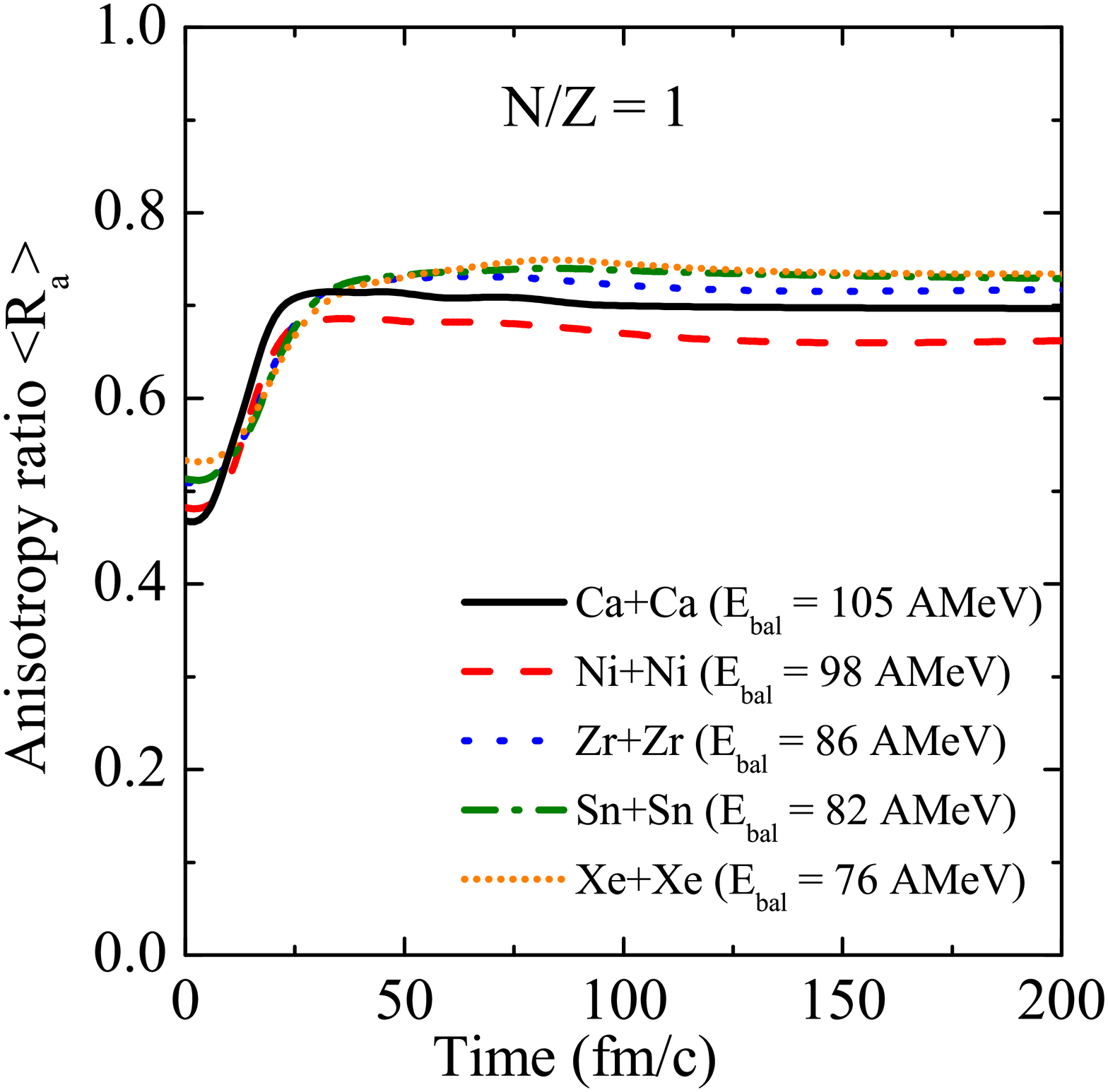}
\caption{ The time evolution of anisotropy ratio for various
systems having N/Z = 1.0.}\label{fig1}
\end{figure}

\par
An anisotropy ratio is an indicator of the global equilibrium of
the system. This represents the equilibrium of the whole system
and does not depend on the local positions. The full global
equilibrium averaged over large number of events will correspond
to $<R_{a}>$ = 1. The $<R_{a}>$ is defined as
\begin{equation}
<R_{a}> =
\frac{\sqrt{p_{x}^{2}}+\sqrt{p_{y}^{2}}}{2\sqrt{p_{z}^{2}}}.
\end{equation}

 From figure 2, we see that anisotropy ratio increases as the reaction proceeds and
finally saturates after the high density phase is over. We also
see that the influence of system size is very less on anisotropy
ratio and hence indicates towards the equilibrium of the system.
Also, the saturation of $<$$R_{a}$$>$ ratio after the high density
phase signifies that the nucleon-nucleon collisions happening
after high density phase do not change the momentum space
significantly.

\section*{Acknowledgments}
 This work has been supported by a grant from Centre of Scientific
and Industrial Research (CSIR), Govt. of India.

\end{document}